\documentclass[runningheads]{llncs}

\usepackage[utf8]{inputenc}
\usepackage{graphicx} 
\usepackage[hyphens]{url}
\usepackage[hidelinks]{hyperref} 

\usepackage{acronym} 
\usepackage{amsfonts} 
\newcommand*\circled[1]{\raisebox{.5pt}{\textcircled{\raisebox{-.9pt} {#1}}}} 
\usepackage[binary-units=true,output-exponent-marker=e]{siunitx}
\usepackage{footmisc}

\usepackage{listings}
\usepackage[dvipsnames]{xcolor}
\lstdefinelanguage{SPARQL}{
	morekeywords={SELECT,STRDT,AS,WHERE,VALUES,BIND,PREFIX},
	sensitive=false,
	morecomment=[l]{####},
	morestring=[b]",
}
\usepackage{fixme}
\fxsetup{
    status=draft,
    author={Note},
    layout=inline, 
    theme=color
}
\usepackage{draftwatermark}
\SetWatermarkAngle{0}
\SetWatermarkVerCenter{2cm}
\SetWatermarkFontSize{9pt}
\SetWatermarkColor{red}
\SetWatermarkText{\fbox{\parbox{.9\paperwidth}{This preprint has not undergone peer review (when applicable) or any post-submission improvements or corrections. The Version of Record of this contribution is published in \textit{The Semantic Web. ESWC 2022. Lecture Notes in Computer Science, vol 13261}, and is available online at \url{https://doi.org/10.1007/978-3-031-06981-9_10}}}}

\newcommand*{\blind}[2]{#1} 
\newcommand*{\blindInvisible}[1]{#1} 

\usepackage{scalerel}
\usepackage{tikz}
\usetikzlibrary{svg.path}

\definecolor{orcidlogocol}{HTML}{A6CE39}
\tikzset{
	orcidlogo/.pic={
		\fill[orcidlogocol]	svg{M256,128c0,70.7-57.3,128-128,128C57.3,256,0,198.7,0,128C0,57.3,57.3,0,128,0C198.7,0,256,57.3,256,128z};
		\fill[white] svg{M86.3,186.2H70.9V79.1h15.4v48.4V186.2z}
		svg{M108.9,79.1h41.6c39.6,0,57,28.3,57,53.6c0,27.5-21.5,53.6-56.8,53.6h-41.8V79.1z
			M124.3,172.4h24.5c34.9,0,42.9-26.5,42.9-39.7c0-21.5-13.7-39.7-43.7-39.7h-23.7V172.4z}
		svg{M88.7,56.8c0,5.5-4.5,10.1-10.1,10.1c-5.6,0-10.1-4.6-10.1-10.1c0-5.6,4.5-10.1,10.1-10.1C84.2,46.7,88.7,51.3,88.7,56.8z};
	}
}

\renewcommand\orcidID[1]{\,\href{https://orcid.org/#1}{\mbox{\scalerel*{
	\begin{tikzpicture}[yscale=-1,transform shape]
	\pic{orcidlogo};
	\end{tikzpicture}
}{|}}}}

\usepackage[backend=biber,
style=numeric,
autocite=inline,
sortcites=true,
sorting=none,
url=true,
isbn=false,
doi=true,
language=english,
urldate=comp,
maxnames=4,
minnames=3,
maxcitenames=1,
mincitenames=1,
giveninits=true,
]{biblatex}
\DeclareFieldFormat{labelnumberwidth}{#1.} 
\DeclareNameAlias{default}{family-given} 

\DeclareSourcemap{\maps[datatype=bibtex]{\map[overwrite]{\step[fieldsource=doi, final]\step[fieldset=url, null]\step[fieldset=eprint, null]}}} 

\begin{document}
\hyphenation{data-type data-types roundTiesToEven ne-ga-tive-Zero}
\acrodef{RDF}{Resource Description Framework}
\acrodef{SPARQL}{SPARQL Protocol And RDF Query Language}
\acrodef{W3C}{World Wide Web Consortium}
\acrodef{XSD}{XML Schema}

\title{Why Not to Use Binary Floating Point Datatypes in RDF}
%
\author{\blind{Jan Martin Keil\orcidID{0000-0002-7733-0193}\and Merle Gänßinger\orcidID{0000-0003-4481-069X}}{Authors}}
%
\institute{\blind{Heinz Nixdorf Chair for Distributed Information Systems,\\
	Institute for Computer Science, Friedrich Schiller University Jena, Jena, Germany\\
	\email{\{jan-martin.keil,merle.gaenssinger\}@uni-jena.de}}{Affiliation}}

\maketitle

\begin{abstract}
The \acs{XSD} binary floating point datatypes are regularly used for precise numeric values in \acs{RDF}.
However, the use of these datatypes for knowledge representation can systematically impair the quality of data and, compared to the \acs{XSD} decimal datatype, increases the probability of data processing producing false results.
We argue why in most cases the \acs{XSD} decimal datatype is better suited to represent numeric values in \acs{RDF}.
A survey of the actual usage of datatypes on the relevant subset of the September 2020 Web Data Commons dataset, containing \num{14778325375} literals from real web data, substantiates the practical relevancy of the described problems:
\SI{29}{\%}--\SI{68}{\%} of binary floating point values are distorted due to the datatype.

\hypersetup{pdfkeywords={Data Quality, Datatypes, Floating Point Precision, Knowledge Graphs, Numerical Stability, RDF, XSD}}
\end{abstract}

\section{Introduction}
\label{sec_introduction}

The \ac{RDF} is the fundamental building block of knowledge graphs and the Semantic Web.
In \ac{RDF}, values are represented as literals.
A literal consists of a lexical form, a datatype, and possibly a language tag.
The use of \ac{XSD} built-in datatypes \cite{W3C2012XSD1.1Datatypes} is recommended by the \ac{RDF} standard \cite[Sec. 5.1]{W3C2014}.
For numeric values, this includes the primitive types \textit{float}, \textit{double} and \textit{decimal} as well as all variations of integer\footnote{\textit{integer}, \textit{long}, \textit{int}, \textit{short}, \textit{byte}, \textit{nonNegativeInteger}, \textit{positiveInteger}, \textit{unsignedLong}, \textit{unsignedInt}, \textit{unsignedShort}, \textit{unsignedByte}, \textit{nonPositiveInteger}, and \textit{negativeInteger}} which are derived from decimal.

The datatype decimal allows to represent numbers with arbitrary precision, whereas the datatypes float and double allow to represent binary floating point values of limited range and precision \cite{W3C2012XSD1.1Datatypes}.
However, in practice, the binary floating point datatypes are regularly used for precise numeric values, although the datatype can not accurately represent these values.
Even a popular ontology guideline~\cite{Noy2001} and a \ac{W3C} working group note~\cite{W3C2006} use binary floating point datatypes for precise numeric values in examples.

In general, binary floating point numbers are meant to approximate decimal values in a fixed length binary representation to limit memory consumption and increase computation speed.
In \ac{RDF}, however, binary floating point numbers are defined to represent the exact value of the binary representation:
Binary floating point values do not approximate typed decimals, as in programming languages, but typed decimals are abbreviations for exact binary floating point values.
This causes ambiguity about the intended meaning of numeric values.
We show that \SI{29}{\%}--\SI{68}{\%} of the floating point values in real web data are distorted due to the datatype.
With regard to the growing use of \ac{RDF} for the representation of data, including research data, this ambiguity is concerning.

Further, the use of binary floating point datatypes for precise numeric values regularly causes rounding errors in the values actually represented, compared to typed values provided as decimals.
Subsequently, error accumulation may significantly falsify the result of processing these values.
Disasters, such as the Patriot Missile Failure, which resulted in 28 deaths, illustrate the potential impact of accumulated errors in real world applications \cite{GAO1992}.
The increasing relevance of knowledge graphs for real-world applications calls for general awareness of these issues in the Semantic Web community.

In this paper, we discuss advantages and disadvantages of the different numeric datatypes.
We demonstrate the practical relevance of outlined problems with a survey of the actual usage of datatypes on the relevant subset of the September 2020 Web Data Commons dataset, containing \num{14778325375} literals from real web data.
We aim to raise awareness of the implications of datatype selection in \ac{RDF} and to enable a more informed choice in the future.
This work is structured as follows:
In \autoref{sec_background}, we give an overview of relevant standards and related work, followed by a comparison of the properties of the binary floating point and decimal datatypes in \autoref{sec_properties}.
In \autoref{sec_discussion}, we discuss the implications of the datatype properties in different use cases.
An approach for automatic problem detection is sketched in \autoref{sec_detection}.
In \autoref{sec_survey}, we present a survey on the use of datatypes in the Word Wide Web that demonstrates the practical relevance of the outlined problems.
Finally, we indicate approaches for the general mitigation of the problems in \autoref{sec_conclusion}.

\section{Background}
\label{sec_background}

Each datatype in \ac{RDF} consists of a lexical space, a value space, and a lexical-to-value mapping.
This is compatible with datatypes in \ac{XSD} \cite{W3C2014}.

\begin{quote}
	The \textbf{value space} of a datatype is the set of values for that datatype \cite{W3C2012XSD1.1Datatypes, W3C2014}.
\end{quote}

\begin{quote}
	The \textbf{lexical space} of a datatype is the prescribed set of strings, which the lexical mapping for that datatype maps to values of that datatype.
	The members of the lexical space are \textbf{lexical representations} (\ac{XSD}) or \textbf{lexical forms} (\ac{RDF}) of the values to which they are mapped \cite{W3C2012XSD1.1Datatypes, W3C2014}.
\end{quote}

\begin{quote}
	The \textbf{lexical mapping} (\ac{XSD}) or \textbf{lexical-to-value mapping} (\ac{RDF}) for a datatype is a prescribed relation which maps from the lexical space of the datatype into its value space \cite{W3C2012XSD1.1Datatypes, W3C2014}.
\end{quote}

\ac{RDF} reuses many of the \ac{XSD} datatypes \cite{W3C2014}.
For non-integer numbers, \ac{XSD} provides the datatypes decimal, float and double.
The XSD datatype \textbf{decimal} (\texttt{xsd:decimal}) represents a subset of the real numbers \cite{W3C2012XSD1.1Datatypes}.

\begin{description}
	\item[Value space of \texttt{xsd:decimal}:] The set of numbers that can be obtained by dividing an integer by a non-negative power of ten: \(\frac{i}{10^n}\) with \(i\in\mathbb{Z}, n\in\mathbb{N}_0, \) precision is not reflected \cite{W3C2012XSD1.1Datatypes}.
	\item[Lexical space of \texttt{xsd:decimal}:] \sloppy The set of all decimal numbers with or without a decimal point \cite{W3C2012XSD1.1Datatypes}.
	\item[Lexical mapping of \texttt{xsd:decimal}:] Set \(i\) according to the decimal digits of the lexical representation and the leading sign, and set \(n\) according to the position of the period or \(0\), if the period is omitted.  If the sign is omitted,``\(+\)'' is assumed \cite{W3C2012XSD1.1Datatypes}.
\end{description}

The \ac{XSD} datatype \textbf{float} (\texttt{xsd:float}) is aligned with the IEEE 32-bit binary floating point datatype \cite{IEEE2008.754}\footnote{As the \ac{XSD} recommendation refers to IEEE 754-2008 version of the standard, we do not refer to the subsequent IEEE 754-2019 version.}, the \ac{XSD} datatype \textbf{double} (\texttt{xsd:double}) is aligned to the IEEE 64-bit binary floating point datatype \cite{IEEE2008.754}. Both represent subsets of the rational numbers. They only differ in their three defining constants \cite{W3C2012XSD1.1Datatypes}.

\begin{description}
	\item[Value space of \texttt{xsd:float} (\texttt{xsd:double}):] The set of the special values  \textit{positiveZero}, \textit{negativeZero}, \textit{positiveInfinity}, \textit{negativeInfinity}, and \textit{notANumber} and the numbers that can be obtained by multiplying an integer \(m\) whose absolute value is less than \(2^{24}\) (double: \(2^{53}\)) with a power of two whose exponent \(e\) is an integer between \(-149\) (double: \(-1074\)) and \(104\) (double: \(971\)): \(m \cdot 2^e\) \cite{W3C2012XSD1.1Datatypes}.
	\item[Lexical space of \texttt{xsd:float} (\texttt{xsd:double}):] The set of all decimal numbers with or without a decimal point, numbers in exponential notation, and the literals \texttt{INF}, \texttt{+INF}, \texttt{-INF}, and \texttt{NaN} \cite{W3C2012XSD1.1Datatypes}.
	\item[Lexical mapping of \texttt{xsd:float} (\texttt{xsd:double}):] Set either the according numeric value (including rounding, if necessary), or the according special value. An implementation might choose between different rounding variants that satisfy the requirements of the IEEE specification.
\end{description}

Numbers with a fractional part of infinite length, like, for example, the rational number \(\frac{1}{3}=0.\bar{3}\) or the irrational number \(\sqrt{2}=1.4142\ldots\), are not in the value space of \texttt{xsd:float} or \texttt{xsd:double}, as a number of finite length multiplied or divided by two is always a number of finite length again.
Consequently, a finite decimal with sufficient precision can exactly represent every possible numeric value or lexical representation of an \texttt{xsd:float} or \texttt{xsd:double}, except of the special values \textit{positiveInfinity}, \textit{negativeInfinity}, and \textit{notANumber}.
In contrast, a finite binary floating point value can not exactly represent every possible decimal value.

Some serialization or query languages for \acs{RDF} provide a shorthand syntax for numeric literals without explicit datatype specification.
In Turtle, TriG and \acs{SPARQL} a number without fraction is an \texttt{xsd:integer}, a number with fraction is an \texttt{xsd:decimal}, and a number in exponential notation is an \texttt{xsd:double} \cite{W3C2014Turtle,W3C2014TriG,W3C2013b}.
In JSON-LD a number without fractions is an \texttt{xsd:integer} and a number with fraction is an \texttt{xsd:double}, to align with the common interpretation of numbers in JSON \cite{W3C2020JSONLD}.
Other \acs{RDF} languages, i.e. RDF/XML, N-Triples, N-Quads, and RDFa, do not provide a shorthand syntax for numeric literals \cite{W3C2014RDFXML,W3C2014NTriples,W3C2014NQuads,W3C2015RDFa}.
Other languages for machine-readable annotation of HTML, which are regularly mapped to RDF, i.e. Microformats\footnote{\url{https://microformats.org}}, and Microdata\footnote{\url{https://html.spec.whatwg.org/multipage/microdata.html}}, do not incorporate explicit datatypes.

In addition to the core \ac{XSD} datatypes, a \ac{W3C} working group note introduces the \texttt{precisionDecimal} datatype \cite{W3C2011a}.
It is aligned to the IEEE decimal floating-point datatypes \cite{IEEE2008.754} and represents a subset of real numbers.
It retains precision and permits the special values \textit{positiveZero}, \textit{negativeZero}, \textit{positiveInfinity}, \textit{negativeInfinity}, and \textit{notANumber}.
Further, it supports exponential notation.
The precision and exponent values of the \texttt{precisionDecimal} datatype are unbounded, but can be restricted in derived datatypes to comply with an actual IEEE decimal floating-point datatype.
However, even though the \ac{RDF} standard permits the use of \texttt{precisionDecimal}, it does not demand its support in compliant implementations \cite{W3C2014}.
Therefore, \ac{RDF} frameworks can not be expected to always support \texttt{precisionDecimal}.

Another \ac{W3C} working group note addresses the selection of proper numeric datatypes for different use cases.
It identified three relevant use cases of numeric values: count, measurement, and constant.
According to the note, the usually appropriate datatypes are (derived datatypes of) the integer datatype for counts, binary floating point datatypes for measurements, and the decimal datatype for constants \cite{W3C2006}.

The popular ontology schema.org\footnote{\url{http://schema.org}, current version 13.0} defines alternative numeric datatypes \texttt{schema:Integer} and \texttt{schema:Float} and their super datatype \texttt{schema:Number}.
A usage note restricts the lexical space of \texttt{schema:Number} to the digits 0 to 9 and at most one full stop.
No further restrictions of the lexical or value space are made.
\texttt{schema:Number} is directly in the range of 91 properties defined by schema.org and \texttt{schema:Integer} is directly in the range of 47 properties.
\texttt{schema:Float} is not directly in the range of any property.

The digital representation or computation of numerical values can cause numerical problems:
An \emph{overflow error} occurs, if a represented value exceeds the maximum positive or negative value in the value space of a datatype.
An \emph{underflow error} occurs, if a represented value is smaller than the minimum positive or negative value different from zero in the value space of a datatype.
A \emph{rounding error} occurs, if a represented value is not in the value space of a datatype and is represented by a nearby value in the value space that is determined by a rounding scheme.
A \emph{cancellation} is caused by the subtraction of nearly equal values and eliminates leading accurate digits. This will expose rounding errors in the input values of the subtraction.
\emph{Error accumulation} is the insidious growth of errors due to the use of a numerically instable sequence of operations.

\section{Properties of Binary Floating Point and Decimal Datatypes in \ac{RDF}}
\label{sec_properties}

Binary floating point and decimal datatypes in the context of \ac{RDF} have individual properties, which make them more or less suitable for specific use cases:

\texttt{xsd:float} and \texttt{xsd:double} permit the use of \textbf{positive and negative infinite values}.
\texttt{xsd:decimal} supports neither positive nor negative infinite values.

\texttt{xsd:float} and \texttt{xsd:double} permit the \textbf{exponential notation}.
Especially in case of numbers with many leading or trailing zeros, this is more convenient and less error-prone to read or write for humans. 
\texttt{xsd:decimal} does not permit the exponential notation.
There is no actual reason for this limitation.
For example, Wikibase\footnote{\label{foot_wikibase}\url{https://wikiba.se/}, \acs{SPARQL} endpoint example: \url{https://query.wikidata.org}} also accepts exponential notation for \texttt{xsd:double}.%
\footnote{Example query: \begin{minipage}[t]{.7\textwidth}\href{https://query.wikidata.org/\#PREFIX\%20xsd\%3A\%20\%3Chttp://www.w3.org/2001/XMLSchema\%23\%3E\%0ASELECT\%20\%28\%221e-9\%22\%5E\%5Exsd\%3Adecimal\%20AS\%20\%3Fd\%29\%20WHERE\%20\%7B\%7D}{\lstinline[language=SPARQL]{PREFIX xsd: <http://www.w3.org/2001/XMLSchema#>}\\\lstinline[language=SPARQL]{SELECT ("1e-9"\^\^xsd:decimal AS ?d) WHERE \{\}}}\end{minipage}}
The \textit{XML Schema Working Group} decided against allowing exponential notation for \texttt{xsd:decimal}, as the requirement to have a decimal datatype permitting exponential notation was already met by \texttt{precisionDecimal} \cite{W3C2006XMLSchemaRQ28}, which, however, has been dropped in the later process of the standardization \cite{W3C2011XSD1.1DatatypesDraft}.

\begin{figure}[tb]
	\centering
	\includegraphics[width=.9\linewidth]{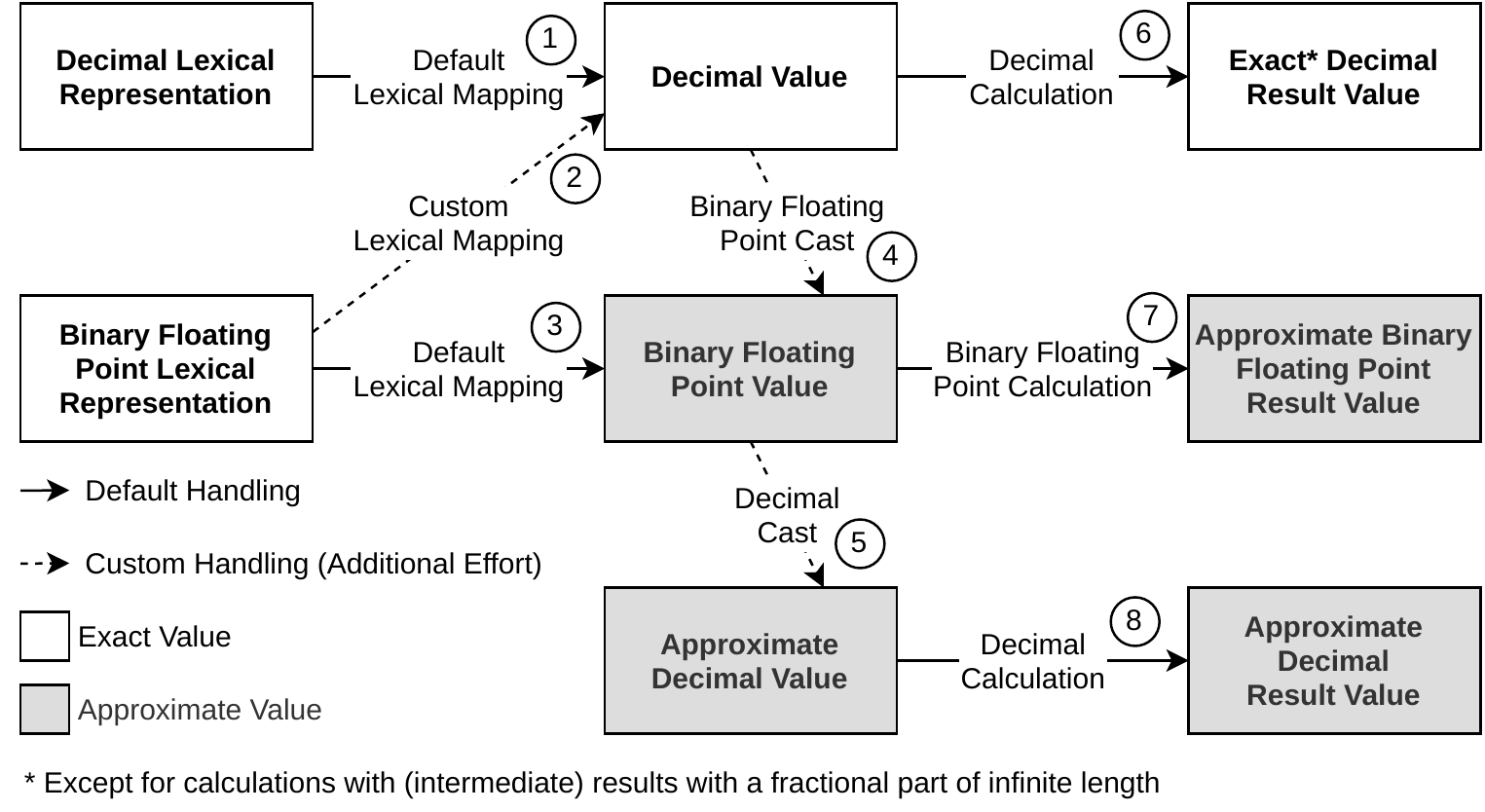}
	\caption{Possible processing paths of numeric literals depending on their datatype.}
	\label{fig_precissiondataflow}
\end{figure}

The value spaces of \texttt{xsd:float} and \texttt{xsd:double} only provide partial \textbf{coverage of the lexical space}.
Therefore, the lexical mapping (\circled{3} in \autoref{fig_precissiondataflow}) might require rounding to a possible binary representation and the actual value might slightly differ from the lexical representation.
For example, \texttt{xsd:float} has no exact binary representation of \(0.1\) and, if using the default \textit{roundTiesToEven} rounding scheme \cite{IEEE2008.754}, the lexical representation \(0.1\) of \texttt{xsd:float} actually maps to a slightly higher binary representation of \(0.10000000149011612\).
Depending on the used \ac{RDF} framework, it might be possible to preserve the exact value of the lexical representation by implementing a custom mapping to decimal (\circled{2} in \autoref{fig_precissiondataflow}).
However, this causes additional development effort and introduces non standard compliant behavior.
The value space of \texttt{xsd:decimal} covers all values in the lexical space.
Therefore, the lexical mapping (\circled{1} in \autoref{fig_precissiondataflow}) always provides the exact numeric value described in the lexical representation without any rounding.
All three datatypes, \texttt{xsd:float}, \texttt{xsd:double}, and \texttt{xsd:decimal}, do not cover the precision reflected by the lexical representation.
The only discussed datatype that preserves precision is \texttt{precisionDecimal}.

The \textbf{accuracy of calculations} based on \texttt{xsd:float} or \texttt{xsd:double} literals (\circled{7} in \autoref{fig_precissiondataflow}) is limited, as a properly implemented RDF framework will use binary floating point arithmetic by default.
For example, this happens during the execution of \acs{SPARQL} queries that include arithmetic functions or aggregations.
Therefore, the calculations might be affected by various numeric problems, i.e. underflow errors, overflow errors, rounding errors, cancellation, and error accumulation.
Calculations based on \texttt{xsd:decimal} literals (\circled{6} in \autoref{fig_precissiondataflow}) will by default use a decimal arithmetic with arbitrary precision.
Therefore, they might only be affected by rounding errors in case of (intermediate) results with a fractional part of infinite length, as well as accumulations of these rounding errors.
This different behavior is demonstrated in \autoref{fig_problems}.
Depending on the used RDF framework, it might be possible to cast between the datatypes (\circled{4} and \circled{5} in \autoref{fig_precissiondataflow}).
However, a value cast from binary floating point to decimal (\circled{5} in \autoref{fig_precissiondataflow}) is still affected by the rounding error of the floating point value caused by the lexical mapping and subsequent calculations (\circled{8} in \autoref{fig_precissiondataflow}) will still result in approximate results only.
In contrast, the results of calculations based on a value cast from decimal to floating point (\circled{4} in \autoref{fig_precissiondataflow}) and based on an initial floating point value (\circled{3} in \autoref{fig_precissiondataflow}) do not differ, if the same rounding method is used.
The SPARQL query in \autoref{fig_problems} and the according result provided by Wikibase\footref{foot_wikibase} demonstrates different numerical problems of the datatypes.
Other SPARQL endpoints, i.e. Virtuoso 8.3\footnote{\url{https://virtuoso.openlinksw.com/}} and Apache Fuseki 5.16.0\footnote{\label{fn_apache_jena}\url{https://jena.apache.org/}}, provide similar results.

\begin{figure}[tb]
	\centering
	\begin{lstlisting}[language=SPARQL]
PREFIX xsd: <http://www.w3.org/2001/XMLSchema#>
SELECT ?datatype
  (xsd:decimal(STRDT("0.1", ?datatype)) AS ?rounded)
  (xsd:decimal(STRDT("1", ?datatype) / STRDT("3", ?datatype)) AS ?roundedInfinit)
  (xsd:decimal(STRDT("1.0000001", ?datatype) - STRDT("1.0000000", ?datatype)) AS ?cancellation)
  (STRDT("1000000000000000000000000000000000000000", ?datatype) AS ?overflow)
  (STRDT("0.0000000000000000000000000000000000000000000001", ?datatype) AS ?underflow)
WHERE {VALUES ?datatype {xsd:float xsd:decimal}}\end{lstlisting}
	
	\begin{tabular}{|l|ll|}
		datatype & xsd:float & xsd:decimal \\
		rounded & \scalebox{1}{0.10000000149011612} & 0.1 \\
		roundedInfinit & \scalebox{1}{0.3333333432674408} & \scalebox{1}{0.33333333333333333333} \\
		cancellation & \scalebox{.825}{0.00000011920928955078125} & 0.0000001 \\
		overflow & Infinity & \scalebox{.825}{1000000000000000000000000000000000000000} \\
		underflow & 0.0 & \scalebox{.825}{0.0000000000000000000000000000000000000000000001} \\
	\end{tabular}

	\caption{Top: A SPARQL query that demonstrates differing numerical problems of the datatypes float and decimal. Bottom: The corresponding query output (transformed), as on \href{https://query.wikidata.org/\#PREFIX\%20xsd\%3A\%20\%3Chttp://www.w3.org/2001/XMLSchema\%23\%3E\%0ASELECT\%20\%3Fdatatype\%0A\%20\%20\%28xsd\%3Adecimal\%28STRDT\%28\%220.1\%22\%2C\%20\%3Fdatatype\%29\%29\%20AS\%20\%3Frounded\%29\%0A\%20\%20\%28xsd\%3Adecimal\%28STRDT\%28\%221\%22\%2C\%20\%3Fdatatype\%29\%20\%2F\%20STRDT\%28\%223\%22\%2C\%20\%3Fdatatype\%29\%29\%20AS\%20\%3FroundedInfinit\%29\%0A\%20\%20\%28xsd\%3Adecimal\%28STRDT\%28\%221.0000001\%22\%2C\%20\%3Fdatatype\%29\%20-\%20STRDT\%28\%221.0000000\%22\%2C\%20\%3Fdatatype\%29\%29\%20AS\%20\%3Fcancellation\%29\%0A\%20\%20\%28STRDT\%28\%221000000000000000000000000000000000000000\%22\%2C\%20\%3Fdatatype\%29\%20AS\%20\%3Foverflow\%29\%0A\%20\%20\%28STRDT\%28\%220.0000000000000000000000000000000000000000000001\%22\%2C\%20\%3Fdatatype\%29\%20AS\%20\%3Funderflow\%29\%0D\%0AWHERE\%20\%7BVALUES\%20\%3Fdatatype\%20\%7Bxsd\%3Afloat\%20xsd\%3Adecimal\%7D\%7D}{\ttfamily http://query.wikidata.org}. }
	\label{fig_problems}
\end{figure}

\section{Implications for the Choice of Numeric Datatypes}
\label{sec_discussion}

The traditional use case of \ac{RDF} is the representation of knowledge.
The XSD floating point datatypes provide two advantages for the knowledge representation compared to \texttt{xsd:decimal}:
Firstly, the permitted representation of positive and negative infinite might be needed in some cases.
Secondly, the exponential notation eases the representation of very large and very small values and reduces the risk of typing errors due to missing or additional zeros.
This would not be an issue in case of proper user interface support.
But popular tools, like, for example, WebProtégé\footnote{\label{fn_protege}\url{https://protege.stanford.edu/}} and Protégé Desktop\footref{fn_protege}, do not help the user here.
Further, projects that manipulate their RDF documents under version control using SPARQL UPDATE queries, custom generation scripts and manual edits do not have such a user interface at all.

However, in most cases, knowledge concerned with numbers deals with exact decimal numbers or intervals of decimal values.
Intervals are typically described with two exact decimal numbers, either with a minimum value and a maximum value (e.g. \([0.05,0.15]\)) or an value and a measurement uncertainty (e.g. \(0.1 \pm 0.05\)) \cite{JCGM2012}.
The binary floating point datatypes do not allow the accurate representation of exactly known or defined numbers in many cases.
In addition, they entail the risk to fool data curators into believing that they stated the exact number, as the lexical representation on first sight appears to be exact.
This becomes even more critical, if \texttt{xsd:double} was used unintentionally due to a shorthand syntax in Turtle, TriG, SPARQL, or JSON-LD.
This way, the use of binary floating point datatypes produces ambiguity in the data:
The intended meaning could either be the actually represented number in the value space or the verbatim interpretation of the lexical representation.
This ambiguity counteracts the basic ideas behind the Semantic Web and linked open data to ease understanding and reuse of data.
Therefore, binary floating point datatypes are not suitable to fulfill the requirements for knowledge representation.

In consequence, the knowledge can not be used for exact calculations without programming overhead.
The possible small rounding errors of binary floating point input values might accumulate to significant errors in calculation results.
Disasters, as the Patriot Missile Failure, illustrate the potential impact of accumulated errors in real world applications \cite{GAO1992}.

This contradicts a \ac{W3C} working group note \cite{W3C2006}, stating that binary floating point datatypes are appropriate for measurements.
It provided the following example representation of a measurement in the interval of \(73.0\) to \(73.2\):
{\begin{verbatim}
_:w eg:value     "73.1"^^xsd:float .
_:w eg:errorRange "0.1"^^xsd:float .
\end{verbatim}}
However, if using the default \textit{roundTiesToEven} rounding scheme \cite{IEEE2008.754}, this example actually represents a measurement in the interval \(72.99999847263098388\) to \(73.19999847561121612\), as \(73.1\) and \(0.1\) are not in the value space of \texttt{xsd:float}.\footnote{Lexical mappings (\textit{roundTiesToEven} rounding scheme): \(73.1 \rightarrow 73.0999984741211\) and \(0.1 \rightarrow 0.10000000149011612\), Interval calculations: \(73.0999984741211 \pm 0.10000000149011612\)}
In consequence, the actual represented error interval misses to cover all points between \(73.19999847561121612\) and \(73.2\).
A common solution for this problem is the use of different rounding schemes for the calculation of the upper and lower bound of the interval (outward rounding) \cite{Neumaier2012}. 
However, this is not provided in current \ac{RDF} frameworks and causes additional programming effort.
The example shows that also in case of measurements binary floating point datatypes have clear disadvantages compared to \texttt{xsd:decimal}.

Further, the use of binary floating point values in \ac{RDF} restricts the selection of the used arithmetic for calculations, as it causes an implementation overhead for the application of decimal arithmetic with arbitrary precision.
It must be mentioned that calculations using decimal arithmetic with arbitrary precision probably are significantly slower, compared to calculations using binary floating point arithmetic with limited precision.
Hence, floating point calculations are better suited for many use cases.
However, in certain cases they are not.
Therefore, the selection of an arithmetic must be up to the application, not to the input data, as applications might widely vary regarding the required accuracy and the numerical conditioning of the underlying problem.

The same problem arises in use cases that involve the comparison of values, like ontology matching or ontology based data validation, because comparison values become blurred due to rounding.
For example, if using the default \textit{roundTiesToEven} rounding scheme, an upper bound of \texttt{"0.1"\^{}\^{}xsd:float} in a constraint still permits a value of \texttt{0.100000001}.
Thus, the use of binary floating point datatypes for knowledge representation can systematically impair the quality of data and increases the probability of data processing producing false results.

In other use cases, \ac{RDF} might be used for the exchange of initially binary floating point values, as computational results or the output of analog-to-digital converters.
If the data to exchange are binary floating point values, the original value can only contain values with an exact binary representation and corruption of data with rounding is impossible.
Thus, the use of floating point datatypes for the exchange of computational results is reasonable.

\section{Automatic Distortion Detection}
\label{sec_detection}

The automatic detection of quality issues is key to an effective quality assurance.
Therefore, \ac{RDF} editors, like Protégé\footref{fn_protege}, or evaluation tools, like the OntOlogy Pitfall Scanner! \cite{Poveda-Villalon2014}, would ideally warn data curators, if the use of binary floating point datatypes would distort numeric values.

A simple test can be implemented by comparing the results of the default mapping to a binary floating point value (\circled{3} in \autoref{fig_precissiondataflow}) followed by an cast to decimal (\circled{5} in \autoref{fig_precissiondataflow}) and a custom mapping to a decimal value (\circled{2} in \autoref{fig_precissiondataflow}).
The SPARQL query in \autoref{fig_detection} demonstrates the approach.

\begin{figure}[h]
	\centering
	\begin{lstlisting}[language=SPARQL]
PREFIX xsd: <http://www.w3.org/2001/XMLSchema#>
SELECT
  (xsd:decimal(?defaultMapping) AS ?float)
  (?customMapping AS ?decimal)
  (xsd:decimal(?defaultMapping) != ?customMapping AS ?distorted)
WHERE {
  VALUES ?lexical {"10" "1" "0.1" "0.5"}
  BIND(STRDT(?lexical, xsd:float) AS ?defaultMapping)
  BIND(STRDT(?lexical, xsd:decimal) AS ?customMapping)
}\end{lstlisting}

	\begin{tabular}{lll}
		\hline 
		float & decimal & distorted \\ 
		\hline 
		0.5 & 0.5 & false \\
		0.10000000149011612 & 0.1 & true \\
		1 & 1 & false \\
		10 & 10 & false \\ 
		\hline 	
	\end{tabular}
	\caption{Top: A SPARQL query that demonstrates an approach to detect number distortion. Bottom: The corresponding query output, as on \href{https://query.wikidata.org/\#PREFIX\%20xsd\%3A\%20\%3Chttp://www.w3.org/2001/XMLSchema\%23\%3E\%0ASELECT\%0A\%20\%20\%28xsd\%3Adecimal\%28\%3FdefaultMapping\%29\%20AS\%20\%3Ffloat\%29\%0A\%20\%20\%28\%3FcustomMapping\%20AS\%20\%3Fdecimal\%29\%0A\%20\%20\%28xsd\%3Adecimal\%28\%3FdefaultMapping\%29\%20\%21\%3D\%20\%3FcustomMapping\%20AS\%20\%3Fdistorted\%29\%0AWHERE\%20\%7B\%0A\%20\%20VALUES\%20\%3Flexical\%20\%7B\%2210\%22\%20\%221\%22\%20\%220.1\%22\%20\%220.5\%22\%7D\%0A\%20\%20BIND\%28STRDT\%28\%3Flexical\%2C\%20xsd\%3Afloat\%29\%20AS\%20\%3FdefaultMapping\%29\%0A\%20\%20BIND\%28STRDT\%28\%3Flexical\%2C\%20xsd\%3Adecimal\%29\%20AS\%20\%3FcustomMapping\%29\%0A\%7D}{\ttfamily http://query.wikidata.org}.}
	\label{fig_detection}
\end{figure}

\section{Datatype Usage Survey}
\label{sec_survey}

To determine the practical relevancy of the described problems, we conducted a survey of the actual usage of datatypes.
The survey is based on the September 2020 edition\footnote{\url{http://webdatacommons.org/structureddata/\#results-2020-1}} of the Web Data Commons dataset \cite{Meusel2014}.
The Web Data Commons dataset provides in several N-Quads files the embedded RDF data of
\num{1.7e9}\,HTML documents extracted from all
\num{3.4e9}\,HTML documents contained in the September 2020 Common Crawl archive.\footnote{\url{https://commoncrawl.org/2020/10/september-2020-crawl-archive-now-available/}}
The September 2020 Web Data Commons dataset is divided into data extracted from embedded JSON-LD, RDFa, Microdata, and several Microformats.
We only considered data from embedded JSON-LD (
\num{7.7e8}\,URLs, 
\num{3.2e10}\,triples) and RDFa (
\num{4.1e8}\,URLs,
\num{5.9e9}\,triples), as Microdata and Microformats do not incorporate explicit datatypes.

We created a Java program based on Apache Jena\footref{fn_apache_jena} to stream and analyze the relevant parts of the Web Data Commons dataset.
The dataset replicates malformed IRIs or literals as they appeared in the original source.
To avoid parsing failures of whole files due to single malformed statements, each line was parsed independently.
These failures were logged in a separate result table.
The main reasons for failures were malformed IRIs and illegal character encodings.
Transaction mechanisms were used to ensure the consistency of the resulting dataset in case of temporary failures of involved systems.
Per source type, dataset file, property, and datatype we measured:

\begin{itemize}
	\item \textbf{UnpreciseRepresentableInDouble}: The number of lexicals that are in the lexical space but not in the value space of \texttt{xsd:double}.
	\item \textbf{UnpreciseRepresentableInFloat}: The number of lexicals that are in the lexical space but not in the value space of \texttt{xsd:float}. 
	\item \textbf{UsedAsDatatype}: The total number of literals with the datatype.
	\item \textbf{UsedAsPropertyRange}: The number of statements that specify the datatype as range of the property.
	\item \textbf{ValidDecimalNotation}: The number of lexicals that represent a number with decimal notation and whose lexical representation is thereby in the lexical space of \texttt{xsd:decimal}, \texttt{xsd:float}, and \texttt{xsd:double}.
	\item \textbf{ValidExponentialNotation}: The number of lexicals that represent a number with exponential notation and whose lexical representation is thereby in the lexical space of \texttt{xsd:float}, and \texttt{xsd:double}.
	\item \textbf{ValidInfOrNaNNotation}: The number of lexicals that equals either \texttt{INF}, \texttt{+INF}, \texttt{-INF} or \texttt{NaN} and whose lexical representation is thereby in the lexical space of \texttt{xsd:float}, and \texttt{xsd:double}.
	\item \textbf{ValidIntegerNotation}: The number of lexicals that represent an integer number and whose lexical representation is thereby in the lexical space of \texttt{xsd:integer}, \texttt{xsd:decimal}, \texttt{xsd:float}, and \texttt{xsd:double}.
\end{itemize}

Unfortunately, the lexical representation of \texttt{xsd:double} literals from embedded JSON-LD was normalized during the creation of the Web Data Commons dataset to always uses exponential notation with one integer digit and up to 16 fractional digits.%
\footnote{\url{https://github.com/jsonld-java/jsonld-java/blob/v0.13.1/core/src/main/java/com/github/jsonldjava/core/RDFDataset.java\#L673}}
This is a legal transformation according to the definition of \texttt{xsd:double}, as the represented value is preserved.
However, this limits the use of the according \textit{Valid…} and \textit{Unprecise…} measures.
At the same time, this demonstrates, that the use of \texttt{xsd:float} or \texttt{xsd:double} might easily cause the loss of information reflected only in the lexical representation by legal transformations during the processing.

The resulting dataset consist of two CSV files containing the measurement results (\num{5.4e7} lines, \SI{0.6}{\gibi\byte} compressed, \SI{11.0}{\gibi\byte} uncompressed) and the parsing failure log (\num{4.5e7} lines, \SI{0.5}{\gibi\byte} compressed, \SI{9.7}{\gibi\byte} uncompressed).
The analysis was conducted with python scripts.
The tool \cite{Gaenssinger2021RdfDatatypeUsageScanner}, the resulting dataset \cite{Keil2021RdfDatatypeUsageDataset}, and the analysis scripts \cite{Keil2021RdfDatatypeUsageAnalysis} are freely available for review and further use under permissive licenses.

For the analysis, we first applied some data cleaning:
Some properties and datatypes where regularly denoted by IRIs in the \texttt{http} scheme as well as in the \texttt{https} scheme.
To enable proper aggregation, the scheme of all IRIs in the dataset were unified to \texttt{http}.
Further, the omission of namespace definitions in the source websites causes the occurrence of prefixed names instead of full IRIs.
All prefixes in datatypes that occurred at least for one datatype more than 1000 times and whose namespace was obvious, have been replaced with the actual namespace.
In the same way, all prefixes in properties that occurred at least for one property more than 1000 times and whose namespace was obvious, have been replaced with the actual namespace.
We did not clean further kinds of typos, like, for example, missing or duplicated \texttt{\#} or \texttt{/} after the namespace.

For the results presentation, we use the following prefixes to abbreviate IRIs:\\
{\footnotesize \begin{tabular}{rl}
	\texttt{dcterms:} & \texttt{http://purl.org/dc/terms/} \\
	\texttt{dv:}      & \texttt{http://rdf.data-vocabulary.org/\#} \\
	\texttt{gr:}      & \texttt{http://purl.org/goodrelations/v1\#} \\
	\texttt{rev:}     & \texttt{http://purl.org/stuff/rev\#} \\
	\texttt{rdf:}     & \texttt{http://www.w3.org/1999/02/22-rdf-syntax-ns\#} \\
	\texttt{schema:}  & \texttt{http://schema.org/} \\
	\texttt{use:}     & \texttt{http://search.yahoo.com/searchmonkey-datatype/use/} \\
	\texttt{vcard:}   & \texttt{http://www.w3.org/2006/vcard/ns\#} \\
	\texttt{xsd:}     & \texttt{http://www.w3.org/2001/XMLSchema\#} \\
\end{tabular}}

\begin{table}[tbh]
	\centering
	\caption{The number of datatype occurrences in the Web Data Commons September 2020 dataset from RDFa and embedded JSON-LD sources in absolute numbers and relative to the total number of literals in the source type. Only the top 10, as well as selected further datatypes are shown.}
	\label{tab_datatype_occurrences}
	\begin{tabular}{l|r}
		\multicolumn{2}{c}{RDFa} \\
		\hline
		Datatype & Occurrences (rel) \\
		\hline
		\texttt{rdf:langString} & \num{3179161585} (.68) \\
		\texttt{xsd:string}     & \num{1305371136} (.28) \\
		\texttt{xsd:dateTime}   &  \num{102987223} (.02) \\
		\texttt{rdf:XMLLiteral} &   \num{62337177} (.01) \\
		\texttt{xsd:integer}    &   \num{21547053} (.00) \\
		\texttt{xsd:float}      &    \num{1025753} (.00) \\
		\texttt{use:sku}        &     \num{729858} (.00) \\
		\texttt{xsd:date}       &     \num{507454} (.00) \\
		\texttt{xsd:boolean}    &     \num{348334} (.00) \\
		\texttt{schema:Date}    &     \num{246995} (.00) \\
		\hline
		\hline
		\texttt{xsd:decimal}    &       \num{8288} (.00) \\
		\texttt{xsd:double}     &        \num{234} (.00) \\
		\texttt{schema:Number}  &          \num{0} (.00) \\
		\texttt{schema:Integer} &          \num{0} (.00) \\
		\texttt{schema:Float}   &          \num{0} (.00) \\
		\hline
	\end{tabular}
	\begin{tabular}{l|r}
		\multicolumn{2}{c}{Embedded JSON-LD} \\
		\hline
		Datatype & Occurrences (rel) \\
		\hline
		\texttt{xsd:string}      & \num{11277500571} (.76) \\
		\texttt{xsd:integer}     &  \num{2021243795} (.14) \\
		\texttt{schema:Date}     &  \num{1313408439} (.09) \\
		\texttt{xsd:double}      &   \num{101959406} (.01) \\
		\texttt{xsd:boolean}     &    \num{26144338} (.00) \\
		\texttt{schema:DateTime} &    \num{25002464} (.00) \\
		\texttt{rdf:langString}  &    \num{12934431} (.00) \\
		\texttt{xsd:float}       &       \num{90895} (.00) \\
		\texttt{xsd:dateTime}    &       \num{12260} (.00) \\
		\texttt{rdf:HTML}        &        \num{5785} (.00) \\
		\hline
		\hline
		\texttt{xsd:decimal}     &           \num{1} (.00) \\
		\texttt{schema:Number}   &           \num{0} (.00) \\
		\texttt{schema:Integer}  &           \num{0} (.00) \\
		\texttt{schema:Float}    &           \num{0} (.00) \\
		\hline
		\multicolumn{2}{c}{}\\
	\end{tabular}
\end{table}

\begin{table}[tbh]
	\centering
	\caption{The number of property occurrences with XSD or schema.org numerical datatypes in the Web Data Commons September 2020 dataset from RDFa and embedded JSON-LD sources in absolute numbers and relative to the total number of numeric literals in the source type. Only the top 10 are shown.}
	\label{tab_numerical_property_occurrences}
	\scalebox{.88}{\begin{tabular}{l|r}
			\multicolumn{2}{c}{RDFa} \\
			\hline
			Property & Occurrences (rel) \\
			\hline
			\texttt{sioc:num\_replies}   & \num{21391187} (.95) \\
			\texttt{gr:hasCurrencyValue} &   \num{525491} (.02) \\
			\texttt{gr:hasMinValue}      &   \num{137018} (.01) \\
			\texttt{gr:amountOfThisGood} &    \num{94978} (.00) \\
			\texttt{gr:hasMaxValue}      &    \num{52772} (.00) \\
			\texttt{vcard:latitude}      &    \num{49428} (.00) \\
			\texttt{vcard:longitude}     &    \num{49428} (.00) \\
			\texttt{gr:hasValue}         &    \num{25800} (.00) \\
			\texttt{dv:count}            &    \num{24672} (.00) \\
			\texttt{dv:price}            &    \num{23936} (.00) \\
			\hline
	\end{tabular}}
	\scalebox{.88}{\begin{tabular}{l|r}
			\multicolumn{2}{c}{Embedded JSON-LD} \\
			\hline
			Property & Occurrences (rel) \\
			\hline
			\texttt{schema:position}             & \num{893910601} (.42) \\
			\texttt{schema:width}                & \num{448036253} (.21) \\
			\texttt{schema:height}               & \num{446308779} (.21) \\
			\texttt{schema:price}                &  \num{71045655} (.03) \\
			\texttt{schema:commentCount}         &  \num{65723049} (.03) \\
			\texttt{schema:ratingValue}          &  \num{26261677} (.01) \\
			\texttt{schema:longitude}            &  \num{17096852} (.01) \\
			\texttt{schema:latitude}             &  \num{17093196} (.01) \\
			\texttt{schema:bestRating}           &  \num{16333042} (.01) \\
			\texttt{schema:userInteractionCount} &  \num{13347182} (.01) \\
			\hline
	\end{tabular}}
\end{table}

Overall, we processed \num{14778325375} literals from embedded JSON-LD and \num{4674734966} literals from RDFa.
\autoref{tab_datatype_occurrences} shows the number of occurrences of the most frequent datatypes.
\autoref{tab_numerical_property_occurrences} shows the most frequently used properties that occurred with numerical datatypes from XSD or schema.org.
Although the use of the schema.org numeric datatypes instead of XSD numeric datatypes is expected by the definition of many schema.org properties, including widely used properties, like, for example, \texttt{schema:position} or \texttt{schema:price}, we did find zero occurrences of schema.org numeric datatypes.
In contrast, the usage of schema.org temporal datatypes \texttt{schema:Date} and \texttt{schema:DateTime} in JSON-LD exceeds the usage of XSD temporal datatypes by orders of magnitude.
The most probable reason for this is the existence of shorthand syntaxes for XSD numeric datatypes. 
This emphasizes the importance of shorthand syntaxes for the choice of datatypes.

As shown in \autoref{tab_datatype_occurrences}, the occurrences of \texttt{xsd:float} in RDFa and \texttt{xsd:double} in embedded JSON-LD surpass the occurrences of \texttt{xsd:decimal} by orders of magnitude.
Remarkably, we did find only one single occurrence\footnote{\url{https://web.archive.org/web/20200919100939/https://open.nrw/dataset/telefonverzeichnis-alphabetisch-oktober-2019-odp}} of \texttt{xsd:decimal} among \num{14778325375} literals from valid triples in embedded JSON-LD sources in the whole Web Data Commons September 2020 dataset.
\autoref{tab_float_property_occurrences} shows properties that most frequently occurred with \texttt{xsd:float} in RDFa and with \texttt{xsd:double} in embedded JSON-LD.
Based on these figures, at least \SI{62}{\%} for \texttt{xsd:float} in RDFa and \SI{54}{\%} for \texttt{xsd:double} in embedded JSON-LD represent (monetary) amounts, position numbers or single rating values, which are not initially binary floating point values.
At least \SI{33}{\%} for \texttt{xsd:float} in RDFa and  \SI{35}{\%} for \texttt{xsd:double} in embedded JSON-LD represent geolocation values, arbitrary quantity values or aggregated values, which might but do not need to origin from initially binary floating point values.
\texttt{rev:rating} and \texttt{schema:ratingValue} cannot be assigned unambiguously to these categories.
This shows, binary floating point numbers are regularly used for not initially binary floating point values.

As expected, because embedded RDF is not the proper place for vocabulary definitions, we found only few cases of property range definitions.
They are limited to 54 unique property-datatype-pairs with two to 153 occurrences and for properties from only five different namespaces.
This does not allow to draw further conclusions.

\begin{table}[tbh]
	\centering
	\caption{The number of property occurrences with \texttt{xsd:float} in RDFa and with \texttt{xsd:double} in embedded JSON-LD in the Web Data Commons September 2020 dataset in absolute numbers and relative to the total number of literals with the same datatype in the same source type. Only the top 10 are shown.}
	\label{tab_float_property_occurrences}
	\scalebox{.95}{\begin{tabular}{l|r}
			\multicolumn{2}{c}{\texttt{xsd:float} in RDFa} \\
			\hline
			Property & Occurrences (rel) \\
			\hline
			\texttt{gr:hasCurrencyValue} &  \num{516256} (.50) \\
			\texttt{gr:hasMinValue} &  \num{134954} (.13) \\
			\texttt{gr:amountOfThisGood} &  \num{94978} (.09) \\
			\texttt{gr:hasMaxValue} &  \num{52772} (.05) \\
			\texttt{vcard:latitude} &  \num{49428} (.05) \\
			\texttt{vcard:longitude} &  \num{49428} (.05) \\
			\texttt{gr:hasValue} &  \num{25800} (.03) \\
			\texttt{dv:price} &  \num{23086} (.02) \\
			\texttt{dv:average} &  \num{21038} (.02) \\
			\texttt{rev:rating} &  \num{20970} (.02) \\
			\hline
	\end{tabular}}
	\scalebox{.95}{\begin{tabular}{l|r}
			\multicolumn{2}{c}{\texttt{xsd:double} in Embedded JSON-LD} \\
			\hline
			Property & Occurrences (rel) \\
			\hline
			\texttt{schema:price} &  \num{49740982} (.49) \\
			\texttt{schema:longitude} &  \num{17055600} (.17) \\
			\texttt{schema:latitude} &  \num{17053362} (.17) \\
			\texttt{schema:ratingValue} &  \num{9928412} (.10) \\
			\texttt{schema:lowPrice} &  \num{2240110} (.02) \\
			\texttt{schema:highPrice} &  \num{1840080} (.02) \\
			\texttt{schema:value} &  \num{1776255} (.02) \\
			\texttt{schema:worstRating} &  \num{311374} (.00) \\
			\texttt{schema:position} &  \num{240577} (.00) \\
			\texttt{schema:minPrice} &  \num{197850} (.00) \\
			\hline
	\end{tabular}}
\end{table}

\begin{table}[tbh]
	\centering
	\caption{The number of numeric notations occurrences in the lexical representation of literals per numeric datatype in the Web Data Commons September 2020 dataset in absolute numbers and relative to the total number of literals with the same datatype. The notation of \texttt{xsd:double} in embedded JSON-LD was normalized during the dataset generation.}
	\label{tab_numeric_notation}
	\begin{tabular}{l|r|r|r|r}
		\multicolumn{5}{c}{Embedded JSON-LD}\\
		\hline
		&\multicolumn{4}{c}{Notation}\\
		\cline{2-5}
		Datatype & Integer & Decimal & Exponential & Inf / NaN \\
		\hline
		\texttt{xsd:decimal} & \num{0} (.00) & \num{1} (\hphantom{.0}1) & \num{0} (.00) & \num{0} (.0) \\
		\texttt{xsd:double} & \num{0} (.00) & \num{0} (.00) & \num{101959382} (\hphantom{.0}1) & \num{24} (.0) \\
		\texttt{xsd:float} & \num{35951} (.40) & \num{24837} (.27) & \num{4252} (.05) & \num{0} (.0) \\
		\texttt{xsd:integer} & \num{2021243613} (\hphantom{.0}1) & \num{0} (.00) & \num{0} (.00) & \num{0} (.0) \\
		\texttt{xsd:long} & \num{36} (\hphantom{.0}1) & \num{0} (.00) & \num{0} (.00) & \num{0} (.0) \\
		\hline
	\end{tabular}
	\begin{tabular}{l|r|r|r|r}
		\multicolumn{5}{c}{RDFa}\\
		\hline
		&\multicolumn{4}{c}{Notation}\\
		\cline{2-5}
		Datatype & Integer & Decimal & Exponential & Inf / NaN \\
		\hline
		\texttt{xsd:decimal} & \num{89} (.01) & \num{7349} (.89) & \num{0} (.00) & \num{0} (.0) \\
		\texttt{xsd:double} & \num{26} (.11) & \num{208} (.89) & \num{0} (.00) & \num{0} (.0) \\
		\texttt{xsd:float} & \num{353851} (.34) & \num{643206} (.63) & \num{0} (.00) & \num{4} (.0) \\
		\texttt{xsd:int} & \num{16751} (.86) & \num{0} (.00) & \num{0} (.00) & \num{0} (.0) \\
		\texttt{xsd:integer} & \num{21507446} (\hphantom{.0}1) & \num{38} (.00) & \num{0} (.00) & \num{0} (.0) \\
		\texttt{xsd:nonNegativeInteger} & \num{585} (\hphantom{.0}1) & \num{0} (.00) & \num{0} (.00) & \num{0} (.0) \\
		\texttt{xsd:positiveInteger} & \num{6} (\hphantom{.0}1) & \num{0} (.00) & \num{0} (.00) & \num{0} (.0) \\
		\hline
	\end{tabular}
\end{table}

\begin{table}
	\centering
	\caption{The number of lexical representation occurrences without exact representation in the value space of \texttt{xsd:float} and \texttt{xsd:double} per numeric datatype in the Web Data Commons September 2020 dataset in absolute numbers and relative to the total number of literals with the same datatype. The notation of \texttt{xsd:double} in embedded JSON-LD was normalized during the dataset generation.}
	\label{tab_unprecise}
	\begin{tabular}{l|r|r|r|r}
		\hline
		& \multicolumn{2}{c}{Embedded JSON-LD} & \multicolumn{2}{|c}{RDFa}\\
		\hline
		& \multicolumn{4}{c}{Unprecise In}\\
		\cline{2-5}
		Datatype & xsd:float & xsd:double & xsd:float & xsd:double\\
		\hline
		\texttt{xsd:decimal} & \num{0} (.00) & \num{0} (.00) & \num{3087} (.37) & \num{3087} (.37) \\
		\texttt{xsd:double} & \num{69648087} (.68) & \num{69646819} (.68) & \num{58} (.25) & \num{58} (.25) \\
		\texttt{xsd:float} & \num{21750} (.24) & \num{21750} (.24) & \num{339583} (.33) & \num{338676} (.33) \\
		\texttt{xsd:int} & - & - & \num{0} (.00) & \num{0} (.00) \\
		\texttt{xsd:integer} & \num{7564635} (.00) & \num{996} (.00) & \num{1492} (.00) & \num{38} (.00) \\
		\texttt{xsd:long} & \num{2} (.06) & \num{0} (.00) & - & - \\
		\texttt{xsd:nonNegativeInteger} & - & - & \num{136} (.23) & \num{0} (.00) \\
		\texttt{xsd:positiveInteger} & - & - & \num{0} (.00) & \num{0} (.00) \\
		\hline
	\end{tabular}
\end{table}

\autoref{tab_numeric_notation} shows the number of occurrences of different notations.
Except of \texttt{xsd:double} in embedded JSON-LD, which is affected by normalization, exponential notation is only little used in the binary floating point datatypes.
Special values occurred only in even more rare cases.
From that, we conclude, that the notation or needed special values are not the crucial consideration behind using binary floating point datatypes.

\autoref{tab_unprecise} shows the number of lexical representations that are not precisely representable in binary floating point datatypes.
\SI{33}{\%} of the represented \texttt{xsd:float} values in RDFa and \SI{24}{\%} in embedded JSON-LD differ from lexical representations.
In embedded JSON-LD the initial lexical representation of \SI{69}{\%} of the \texttt{xsd:double} values must either have contained more then 17 significant digits or already been differing from the represented value.
Referring to the most common properties used with \texttt{xsd:double} in embedded JSON-LD, shown in \autoref{tab_float_property_occurrences}, the frequent occurrence of values with more then 17 significant digits is implausible.
All together, this shows that \SI{29}{\%}--\SI{68}{\%} of the  values with binary floating point datatype in real web data are distorted due to the datatype.

\section{Conclusion}
\label{sec_conclusion}

Binary floating point numbers are meant to approximate decimal values to reduce memory consumption and increase computation speed.
However, in RDF, decimals are used to approximate binary floating point numbers.
This way, the use of binary floating point datatypes in \ac{RDF} produces ambiguity in represented knowledge and restricts the choice of the arithmetic in standards compliant implementations.
Its use can systematically impair the quality of data and falsify the results of data processing.
This can cause serious impacts in real world applications.

A radical solution to remove ambiguity and restrictions on the choice of the arithmetic in standard compliant implementations, and that does not require to update existing data, would be the deprecation and replacement of \texttt{xsd:float} and \texttt{xsd:double} with an extended mandatory \texttt{xsd:decimal} datatype in \ac{RDF}.
The extended \texttt{xsd:decimal} datatype should additionally permit exponential notation and the special values \textit{positiveInfinity}, \textit{negativeInfinity}, and \textit{notANumber}, to cover the whole lexical space and value space of \texttt{xsd:float} and \texttt{xsd:double}.
It should also become the default datatype in the different serialization and query languages for numbers in decimal and exponential notation and be used for interpretation instead of the old deprecated datatypes, if these are used in existing data.
One or several additional new datatypes with hexadecimal lexical representations should be used for the actual representation of binary floating point values.
However, this radical solution would once make a decision for existing data in favor of the verbatim interpretation of the lexical representation.

A more cautious mitigation of the problem should tackle the disadvantages of \texttt{xsd:decimal}:
It would be desirable to introduce in \ac{RDF} mandatory support for
(a) an exponential notation for the decimal datatype, and
(b) a decimal datatype that supports infinite values, like \texttt{precisionDecimal},
to eliminate these disadvantages.
Further, binary floating point datatypes should only be used for numeric values if
(a) a representation of infinity is required, or
(b) the original source provides binary floating point values.
In general, \texttt{xsd:decimal} must become the first choice for the representation of numbers.
Semantic Web teaching materials should clearly name the disadvantages of the binary floating point datatypes, shorthand syntaxes should in future prioritize the decimal datatype, and Semantic Web tools should hint to use \texttt{xsd:decimal}.

\blindInvisible{\subsubsection{Acknowledgments.}
Many thanks to Alsayed Algergawy, Sheeba Samuel, Sirko Schindler, Eberhard Zehendner, and the first author's supervisor Birgitta König-Ries, as well as 5 anonymous reviewers for very helpful comments on earlier drafts of this manuscript.}

\blindInvisible{\subsubsection{Author Contributions.}
Study conception and design, analysis and interpretation of results, and draft manuscript preparation were performed by Jan Martin Keil.
Data collection was performed by Merle Gänßinger and Jan Martin Keil.
All authors read and approved the final manuscript.}

\phantomsection
\addcontentsline{toc}{section}{\refname}
\sloppy
\printbibliography

\end{document}